\documentclass[intlimits,twoside,a4paper]{article}

\usepackage{amsmath,amssymb}
\usepackage{graphicx}
\usepackage{wrapfig}

\usepackage[T2A]{fontenc}
\usepackage[cp1251]{inputenc}

\usepackage[eqsecnum]{cmpj2}

\issue{2012}{15}{4}{43005}

\doinumber{10.5488/CMP.15.43005}

\title[Goldstone mode singularities in $O(n)$ models]{Goldstone mode singularities in $O(n)$ models}

\author[J. Kaupu\v{z}s, R.V.N. Melnik, J. Rim\v{s}\=ans]
{J. Kaupu\v{z}s\refaddr{label1,label2},
R.V.N. Melnik\refaddr{label3},
J. Rim\v{s}\=ans\refaddr{label1,label2}
}
\addresses{
\addr{label1} Institute of Mathematics and Computer Science,
University of Latvia, \\ 29 Rai\c{n}a Boulevard, LV--1459 Riga, Latvia
\addr{label2} Institute of Mathematical Sciences and  Information Technologies,
University of Liepaja, \\ 14 Liela St., Liepaja LV--3401, Latvia
\addr{label3} Wilfrid Laurier University, Waterloo, Ontario, Canada, N2L 3C5
}

\authorcopyright{J. Kaupu\v{z}s, R.V.N. Melnik, J. Rim\v{s}\=ans, 2012}

\date{Received July 3, 2012, in final form August 28, 2012}

\begin{document}

\maketitle

\begin{abstract}
Monte Carlo (MC) analysis of the Goldstone mode singularities
for the transverse and the longitudinal correlation functions, behaving as
$G_{\perp} ({\bf k}) \simeq a k^{-\lambda_{\perp}}$
and $G_{\parallel} ({\bf k}) \simeq b k^{-\lambda_{\parallel}}$ in the ordered
phase at $k \to 0$, is performed in the three-dimensional $O(n)$ models with $n=2,4,10$.
Our aim is to test some challenging theoretical predictions, according to which
the exponents $\lambda_{\perp}$ and $\lambda_{\parallel}$ are non-trivial
($3/2<\lambda_{\perp}<2$ and $0<\lambda_{\parallel}<1$ in three dimensions)
and the ratio $b M^2/a^2$ (where $M$ is a spontaneous magnetization) is universal.
The trivial standard-theoretical values are $\lambda_{\perp}=2$ and
$\lambda_{\parallel}=1$. Our earlier MC analysis gives $\lambda_{\perp}=1.955 \pm 0.020$
and $\lambda_{\parallel}$ about $0.9$ for the $O(4)$ model.
A recent MC estimation of $\lambda_{\parallel}$, assuming
corrections to scaling of the standard theory, yields $\lambda_{\parallel} = 0.69 \pm 0.10$
for the $O(2)$ model. Currently, we have performed a similar MC estimation
for the $O(10)$ model, yielding $\lambda_{\perp} = 1.9723(90)$.
We have observed that the plot of the effective transverse exponent for the $O(4)$ model
is systematically shifted down with respect to the same plot for the $O(10)$ model
by $\Delta \lambda_{\perp} = 0.0121(52)$. It is consistent with the idea that
$2-\lambda_{\perp}$ decreases for large $n$ and tends to zero at $n \to \infty$.
We have also verified and confirmed the expected universality of
$b M^2/a^2$ for the $O(4)$ model, where simulations at two different
temperatures (couplings) have been performed.
 \\ Goldstone mode singularities
\pacs 05.10.Ln, 
      75.10.Hk, 
      05.50.+q 

\end{abstract}

\section{Introduction}
\label{intro}

Our work is devoted to the Monte Carlo (MC) investigation of the Goldstone mode effects in $n$-com\-ponent
vector-spin models ($O(n)$ models), which have $O(n)$ global rotational symmetry
at zero external field ${\bf h}$. The Hamiltonian $\mathcal{H}$ is given by
\begin{equation}
\frac{\mathcal{H}}{T}=-\beta \left( \sum\limits_{\langle i j \rangle}
{\bf s}_i {\bf s}_j + \sum_i {\bf h s}_i \right) \;,
\end{equation}
where $T$ is temperature, ${\bf s}_i \equiv {\bf s}({\bf x}_i)$ is the
$n$--component vector of unit
length, i.~e., the spin variable of the $i$-th lattice site with coordinate ${\bf x}_i$,
and $\beta$ is the coupling constant. The summation takes place over
all nearest neighbors in the lattice with periodic boundary conditions.

The Fourier-transformed longitudinal and transverse correlation functions are
\begin{eqnarray}
G_{\parallel}({\bf k}) &=& N^{-1} \sum\limits_{\bf x} \tilde G_{\parallel}({\bf x}) \re^{-\ri{\bf kx}}\;,
\label{eq:GFpar} \\
G_{\perp}({\bf k}) &=& N^{-1} \sum\limits_{\bf x} \tilde G_{\perp}({\bf x}) \re^{-\ri{\bf kx}}
\label{eq:GFperp} \;,
\end{eqnarray}
where $\tilde G_{\parallel}({\bf x})$ and $\tilde G_{\perp}({\bf x})$ are the corresponding
two-point correlation functions in the coordinate space.

In the thermodynamic limit below the critical temperature (at $\beta>\beta_\mathrm{c}$),
the magnetization $M(h)$ and the
correlation functions exhibit Goldstone mode power-law singularities:
\begin{eqnarray}
& M(h) - M(+0) \propto h^{\rho} & \text{at} \qquad h \to 0 \;, \label{M} \\
& G_{\perp}({\bf k}) = a \, k^{-\lambda_{\perp}} & \text{at} \qquad h=+0 \;\; \mbox{and} \;\; k \to 0 \;, \label{a} \\
& G_{\parallel}({\bf k}) = b \, k^{-\lambda_{\parallel}} & \text{at} \qquad h=+0 \;\; \mbox{and} \;\; k \to 0 \label{b} \;,
\end{eqnarray}
where $a$ and $b$ are the amplitudes.

There exist different theoretical predictions for the values of the exponents in these expressions.
In a series of theoretical works (e.~g.,~\cite{Law1,Law2,HL,Tu,SH78,ABDS99,Dupuis}), it has been claimed
that these exponents are exactly $\rho =1/2$ at $d=3$, $\lambda_{\perp}=2$ and $\lambda_{\parallel}=4-d$.
Here, $d$ is the spatial dimensionality $2 < d < 4$. These theoretical approaches are further
referred to as the standard theory.

More non-trivial universal values are expected according to~\cite{K2010}, such that
\begin{eqnarray}
&&d/2 < \lambda_{\perp} < 2 \;, \label{eq:pred1} \\
&&\lambda_{\parallel} = 2 \lambda_{\perp} - d \;, \label{eq:pred2} \\
&&\rho = (d/\lambda_{\perp})-1  \label{eq:pred3}
\end{eqnarray}
hold for $2<d<4$. These relations were obtained in~\cite{K2010} by analyzing self-consistent
diagram equations for correlation functions without cutting the perturbation series.
As introduced in~\cite{MC_Ising,K2012}, we will call this approach the GFD (grouping of Feynman diagrams)
theory. Apart from the mathematical analysis, certain physical arguments were also provided~\cite{K2010}
to show that $\lambda_{\perp}=2$ could not be the correct result for the $XY$ model ($n=2$) within $2<d<4$.

Several MC simulations were performed in the past~\cite{DHNN,EM,EHMS,EV} to verify the compatibility of MC data
with some standard--theoretical expressions, where the exponents are fixed.
In recent years, we performed a series of accurate MC simulations~\cite{KMR08,KMR10,K2012}
for remarkably larger lattices than previously with an aim to reexamine the theoretical predictions
by evaluating the exponents in~(\ref{eq:pred1})--(\ref{eq:pred3}).
In particular, lattices of the linear sizes $L \leqslant 512$ for $n=2$ and $L \leqslant 350$ for $n=4$
were simulated in our papers~\cite{KMR08,K2012} and~\cite{KMR10}, respectively. These $L$ values
remarkably exceed the largest sizes simulated by other authors, i.~e., $L=160$ for $n=2$ in~\cite{EHMS}
and $L=120$ for $n=4$ in~\cite{EM,EV}. In the current work, the $O(10)$ model is simulated up to $L = 384$.

The relations~(\ref{eq:pred1}) and~(\ref{eq:pred2}) are
consistent with MC simulation results for the 3D $O(4)$ model~\cite{KMR10}, where an estimate
$\lambda_{\perp}=1.955 \pm 0.020$ was found.
It was  also stated that the behavior of the longitudinal correlation function
is well consistent with $\lambda_{\parallel}$ about $0.9$ rather than with the
standard-theoretical value $\lambda_{\parallel}=1$.
According to~(\ref{eq:pred3}), we have $1/2 < \rho <1$ in three dimensions.
It is consistent with the MC estimate $\rho = 0.555(17)$ for the 3D $XY$ model~\cite{KMR08},
which corresponds to $\lambda_{\perp} = 1.929(21)$ according to~(\ref{eq:pred3}).
A clear MC evidence that the behavior of $G_{\parallel}({\bf k})$  is not quite consistent
with the standard--theoretical predictions has been recently provided~\cite{K2012},
where an estimate $\lambda_{\parallel} = 0.69 \pm 0.10$ has been obtained for the 3D $XY$
(i.~e., 3D $O(2)$) model (at $\beta=0.55$), assuming corrections to scaling of the standard theory.

In the actual study, we have extended our MC simulations and analysis to include the $n=10$
case and to test the $n$-dependence of the exponents. Apart from the exponents,
we have performed here an extended analysis of the $O(4)$ model to verify the expected
universality of the ratio $b M^2/a^2$~\cite{K2010}, where $M \equiv M(+0)$ is the spontaneous magnetization,
$a$ and $b$ are the amplitudes in~(\ref{a}) and~(\ref{b}).

\section{Simulation results}
\label{sec:simu}

\begin{figure}[!t]
\begin{center}
\includegraphics[width=0.6\textwidth]{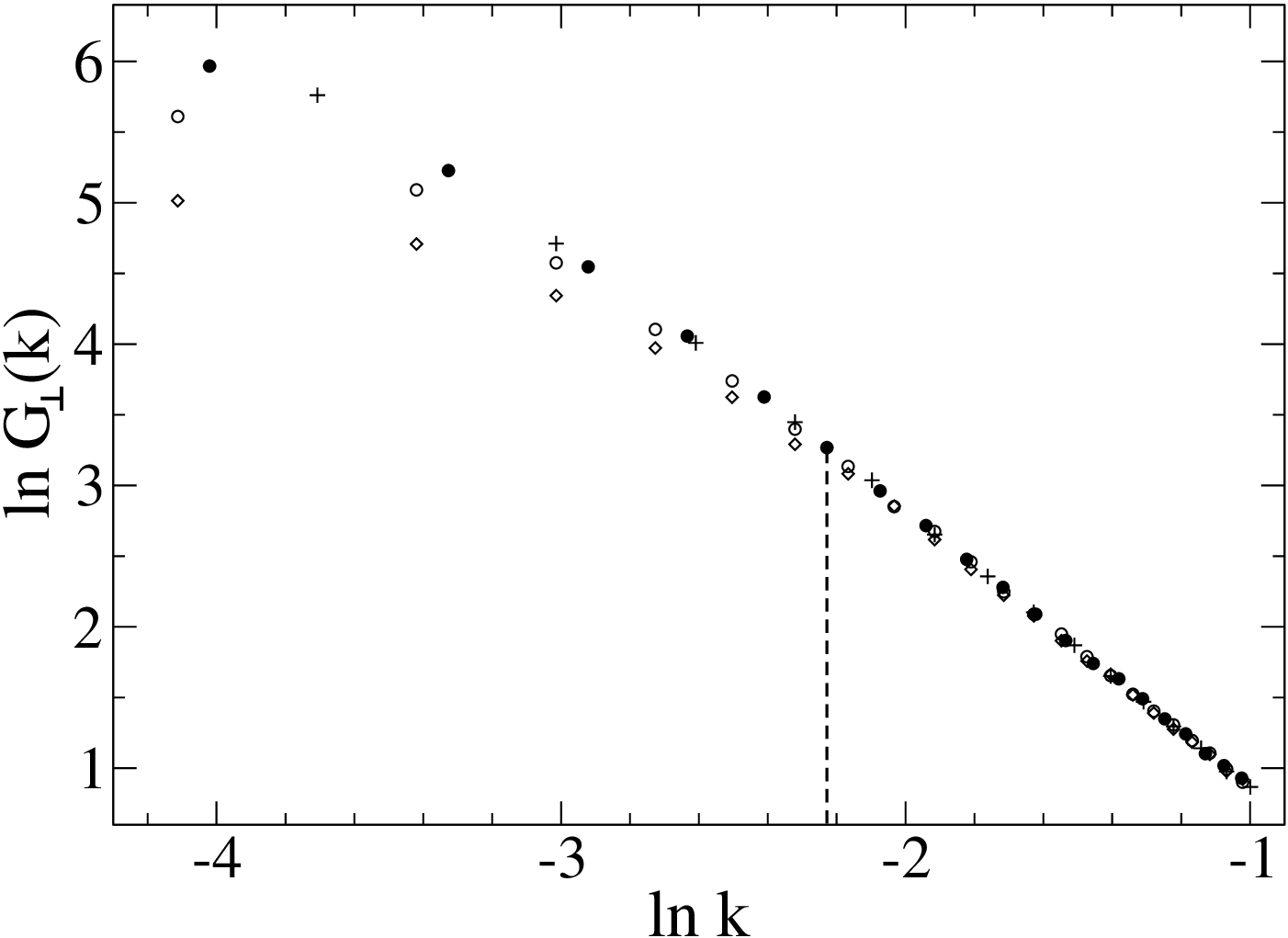}
\end{center}
\caption{Log-log plots of the transverse correlation function
$G_{\perp} ({\bf k})$ at $h=h_{\mathrm{min}}=0.00021875$ and $L=350$ (solid circles),
$h=h_{\mathrm{min}}$ and $L=256$ (pluses), $h=2h_{\mathrm{min}}$ and $L=384$ (empty circles),
as well as at $h=4h_{\mathrm{min}}$ and $L=384$ (empty diamonds). Statistical errors are about
the symbol size or smaller. The lower value $k^*$ of the wave vector magnitude, used in estimations of the
exponent $\lambda_{\perp}$, is indicated by a vertical dashed line.}
\label{transv}
\end{figure}

\begin{figure}[!b]
\begin{center}
\includegraphics[width=0.6\textwidth]{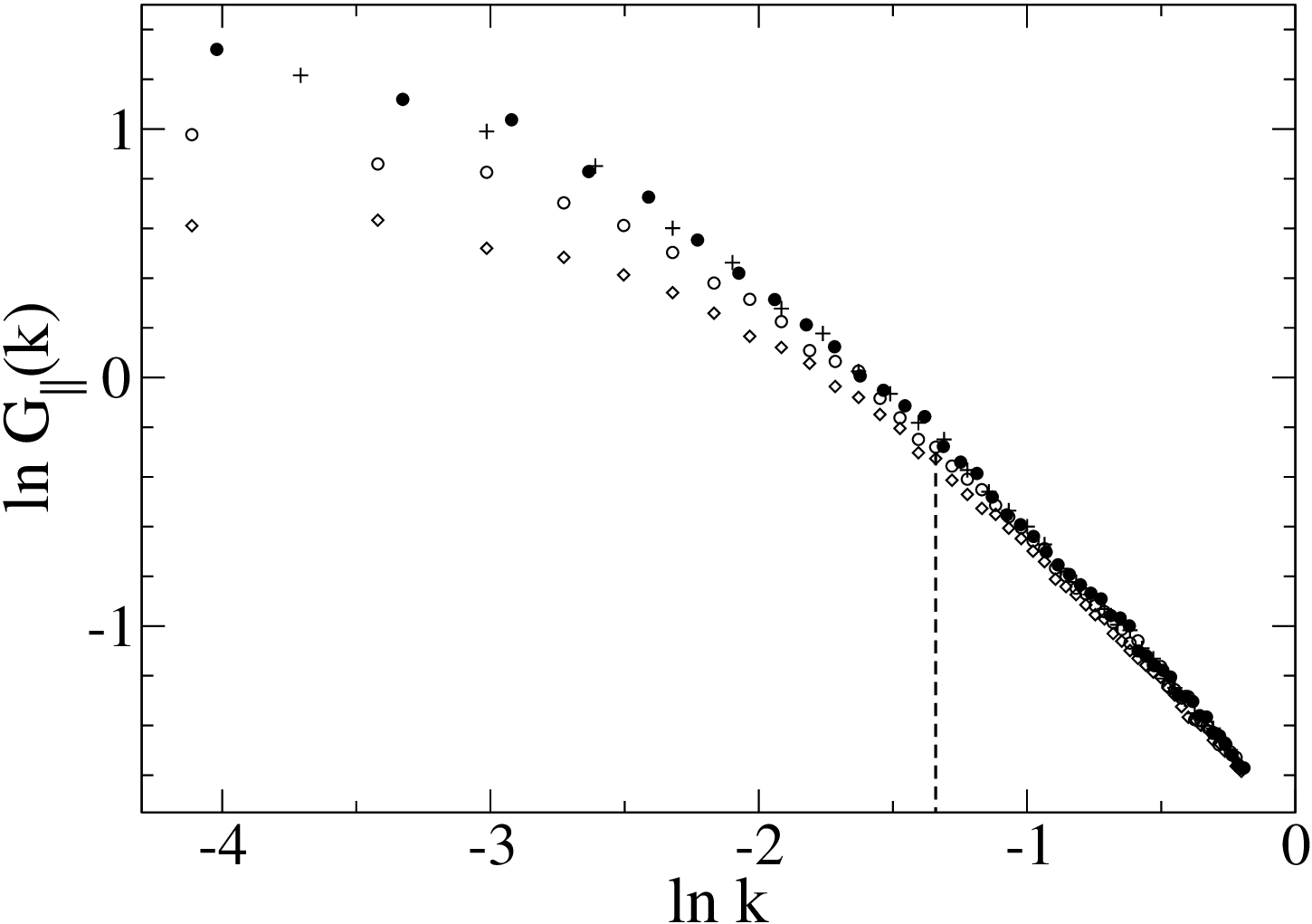}
\end{center}
\caption{Log-log plots of the longitudinal correlation function
$G_{\parallel} ({\bf k})$ at $h=h_{\mathrm{min}}=0.00021875$ and $L=350$ (solid circles),
$h=h_{\mathrm{min}}$ and $L=256$ (pluses), $h=2h_{\mathrm{min}}$ and $L=384$ (empty circles),
as well as at $h=4h_{\mathrm{min}}$ and $L=384$ (empty diamonds). Statistical errors are about
the symbol size. The lower value $k^*$ of the wave vector magnitude, used in estimations of the
exponent $\lambda_{\parallel}$, is indicated by a vertical dashed line.}
\label{long}
\end{figure}

We simulated the 3D $O(10)$ model by a modified Wolff cluster algorithm, used also in~\cite{KMR08,KMR10}, and evaluated the Fourier-transformed correlation functions by techniques described in~\cite{KMR10}.
The standard Wolff cluster algorithm~\cite{Wolff} was modified to enable simulations at nonzero
external field $\mathbf{h}$.
Simple cubic lattices of the linear size up to $L=384$ were simulated
at $\beta = 3$ and $h= \mid \mathbf{h} \mid = h_{\mathrm{min}}, \ 2h_{\mathrm{min}}, \ 4h_{\mathrm{min}}$,
where $h_{\mathrm{min}} = 0.00021875$.
The coupling constant $\beta=3$ corresponds to the ordered phase, since the spontaneous magnetization $M(+0)$ is about
$0.467$ in this case~--- see section~\ref{mag} for details. This value of $M(+0)$ is comparable with those
for the $O(2)$ and $O(4)$ models in our previous MC simulations~\cite{KMR08,KMR10}.
The simulation results for the correlation functions $G_{\perp} ({\bf k})$ and $G_{\parallel} ({\bf k})$
in the $\langle 100 \rangle$ crystallographic direction at the three values of $h$
and different sizes $L$ are illustrated in figures~\ref{transv} and~\ref{long}.
It is important for an estimation of the exponents $\lambda_{\perp}$ and $\lambda_{\parallel}$
to ensure that the finite-size as well as finite-$h$ effects are small. This condition
is satisfied for $k>k^*$, where the values of $k^*$ are indicated in the figures by vertical
dashed lines.

\section{Estimation of the exponents}
\label{expo}

\begin{figure}[!b]
\begin{center}
\includegraphics[width=0.6\textwidth]{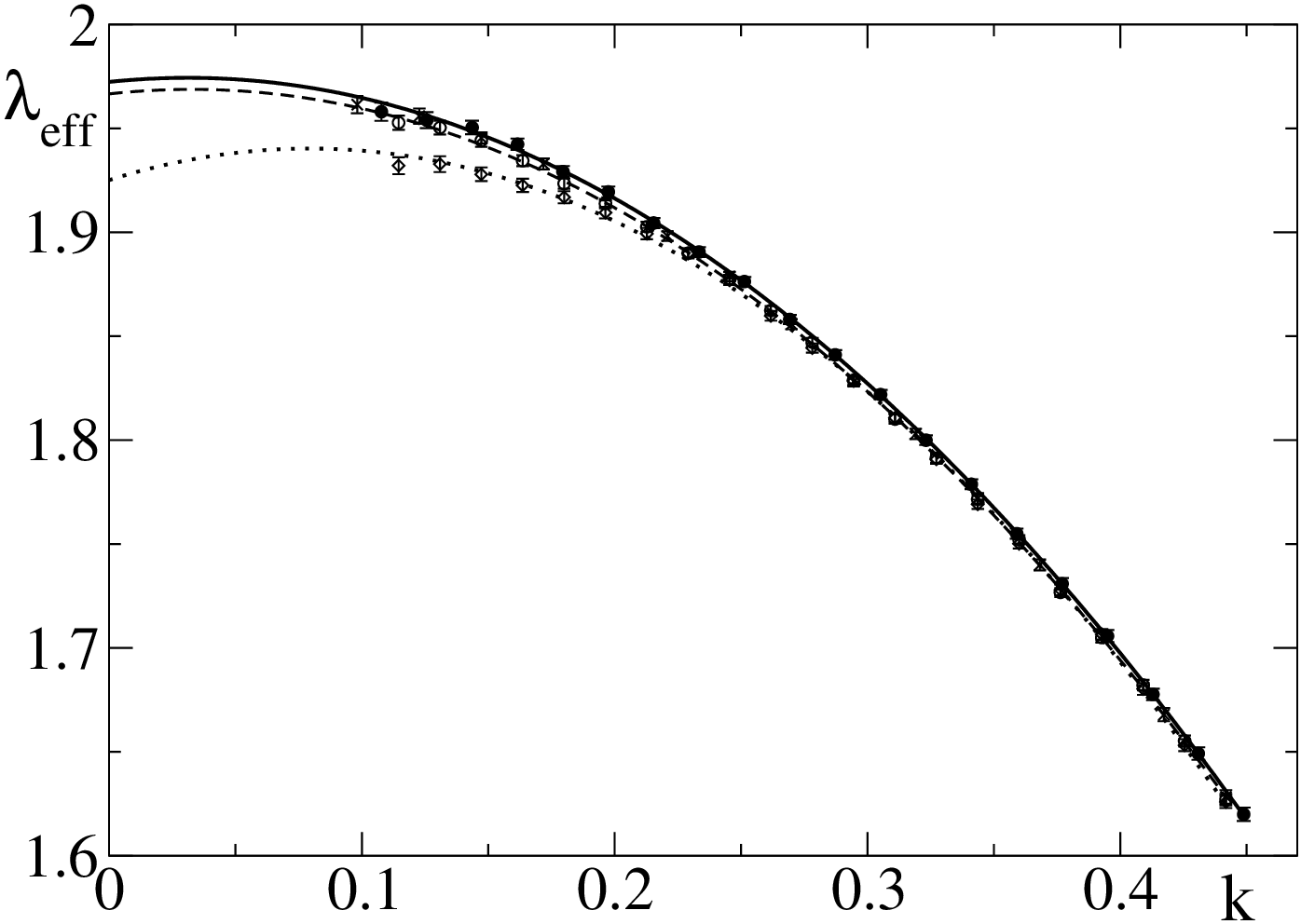}
\end{center}
\caption{The transverse effective exponent
$\lambda_{\mathrm{eff}}(k)$, evaluated from the
$\ln G_{\perp} ({\bf k})$ vs $\ln k$ fits within $[k,6k]$ at $h=h_{\mathrm{min}}=0.00021875$ and $L=350$ (solid circles),
$h=h_{\mathrm{min}}$ and $L=256$ (exes), $h=2h_{\mathrm{min}}$ and $L=384$ (empty circles),
as well as at $h=4h_{\mathrm{min}}$ and $L=384$ (empty diamonds). The fits to~(\ref{eq:fitansatz}) of the largest-$L$ data
at $h=h_{\mathrm{min}}$, $h=2h_{\mathrm{min}}$ and $h=4h_{\mathrm{min}}$ are shown by solid, dashed and dotted lines,
respectively.}
\label{transvexpo}
\end{figure}

Here we estimate the exponents $\lambda_{\perp}$ and $\lambda_{\parallel}$, describing
the behavior of the correlation functions in the limit
${k \to 0}$, ${h \to 0}$, ${L \to \infty}$, taking the limit ${L \to \infty}$ at first, followed by ${h \to 0}$.
For this purpose, first we find good approximations of the effective exponents at
${h \to 0}$, ${L \to \infty}$,
and then fit these $k$-dependent effective exponents to evaluate their asymptotic
values at $k \to 0$. By comparing the simulation results for different $L$ and $h$, we conclude
that the largest-$L$ and smallest-$h$ data for $k > k^*$ with a good enough accuracy correspond to the
thermodynamic limit at $h=+0$, i.~e., ${h \to 0}$, ${L \to \infty}$.
We have tested this precisely by looking how the estimates of the effective exponents depend on $L$ and $h$.
This method of analysis was applied in~\cite{KMR10,K2012}. The effective transverse exponent
$\lambda_{\mathrm{eff}}(k)$ for the $O(4)$ model was evaluated in~\cite{KMR10} from the slope of the
$\ln G_{\perp} ({\bf k})$ vs $\ln k$ plot within $[k,4k]$. Here we use a wider interval~--- $[k,6k]$,
because we have found that the $\lambda_{\mathrm{eff}}(k)$ data in this case can be perfectly fit by a parabola
\begin{equation}
\lambda_{\mathrm{eff}}(k) = \lambda_{\perp} + a_1 k + a_2 k^2 \;,
\label{eq:fitansatz}
\end{equation}
the finite-size and finite-$h$ effects being very small.
The ansatz~(\ref{eq:fitansatz}) is consistent with the general statement
$\lim_{k \to 0} \lambda_{\mathrm{eff}}(k) = \lambda_{\perp}$ (in the thermodynamic limit at $h=+0$)
and with corrections to scaling of the standard theory, where the correlation
functions are supplied with correction factors in the form of an expansion in powers
of $k^{4-d}$ and $k^{d-2}$~\cite{SH78,Law1}.
Some of the fit results are shown in figure~\ref{transvexpo}.
We have performed a series of fits at different sizes $L$ for the smallest-$h$ value $h=h_{\mathrm{min}}=0.00021875$.
At the largest size $L=350$ for this $h$, the effective exponent $\lambda_{\mathrm{eff}}(k)$ was fit
within $k \in [k_5,k_{25}]$, where $k_{\ell} = 2 \pi \ell/350$ are the possible discrete values of $k$.
Similar fit intervals were chosen
for all $L$. These fits to~(\ref{eq:fitansatz}) give us $\lambda_{\perp} = 1.9680(84)$ at $L=128$,
$\lambda_{\perp} = 1.9840(98)$ at $L=192$, $\lambda_{\perp} = 1.9727(86)$ at $L=256$ and
$\lambda_{\perp} = 1.9723(90)$ at $L=350$. As we can see, the finite-size effects are smaller than the
statistical error bars. The $\lambda_{\mathrm{eff}}(k)$ data for $L=350$ and $L=256$ at $h=h_{\mathrm{min}}$
are shown in figure~\ref{transvexpo} by solid circles and exes, respectively. The corresponding fit curves
lie practically on top of each other. Therefore, only that one for $L=350$ is shown by solid line.
The fit curves for $h=2 h_{\mathrm{min}}$ and $h=4h_{\mathrm{min}}$ at $L=384$ (the largest size)
are also depicted here to see the finite-$h$ effects. These fits give us $\lambda_{\perp}=1.9251(86)$
at $h=4 h_{\mathrm{min}}$ and $\lambda_{\perp}=1.9666(91)$ at $h=2 h_{\mathrm{min}}$.
A rather fast convergence to the $h=+0$ limit is evident. According to this discussion,
the fit result $\lambda_{\perp} = 1.9723(90)$, obtained at $h=h_{\mathrm{min}}$
and $L=350$, with a good accuracy corresponds to the thermodynamic limit at $h=+0$.
Besides, the systematical errors due to finite-size and finite-$h$ effects  are probably smaller than
the statistical error bars.

Another possible source of systematical errors is the existence of
non-trivial corrections to scaling, which are not included in~(\ref{eq:fitansatz}).
These are corrections to scaling of the GFD theory~\cite{K2010},
represented by an expansion in powers of $k^{2-\lambda_{\perp}}$,
$k^{\lambda_{\perp} - \lambda_{\parallel}}$ and $k^{\lambda_{\parallel}}$. Nevertheless, the actual estimation,
where only the standard-theoretical corrections have been included, is well justified as a test of consistency of
the standard theory. The existence of a small correction-to-scaling exponent $2 - \lambda_{\perp}$ can make
the extrapolation of the $\lambda_{\mathrm{eff}}(k)$ plots unreliable. However, since the
$\lambda_{\mathrm{eff}}(k)$ data are really well described by a parabola, it might be true that
the amplitude of such a correction term is small and the estimate $\lambda_{\perp} = 1.9723(90)$ is quite reasonable.
In any case, this estimation shows a small deviation from the standard-theoretical picture, where
(\ref{eq:fitansatz}) should hold at small enough $k$ with $\lambda_{\perp} = 2$.
This deviation can be indeed small at $n=10$, since $\lambda_{\perp} \to 2$ is expected
in the limit $n \to \infty$, corresponding to the known behavior of the spherical model~\cite{PeTo}.

We have also attempted to evaluate the longitudinal exponent $\lambda_{\parallel}$ from the
$G_{\parallel}(\mathbf{k})$ data within $k>k^*$, where $k^*$ is indicated in figure~\ref{long}
by a vertical dashed line. We have found that the longitudinal effective exponent, extracted
from the data within $[k,4k]$, can be perfectly approximated by a parabola. It leads to an estimate
$\lambda_{\parallel} = 0.85 \pm 0.06$. The error bars indicated here  include a statistical standard error
as well as a systematical error due to finite-$h$ effects. However, due to a rather large extrapolation gap
(from $\approx 1.17$ to $\approx 0.85$), we consider this estimation as a preliminary one.

\begin{figure}[ht]
\begin{center}
\includegraphics[width=0.6\textwidth]{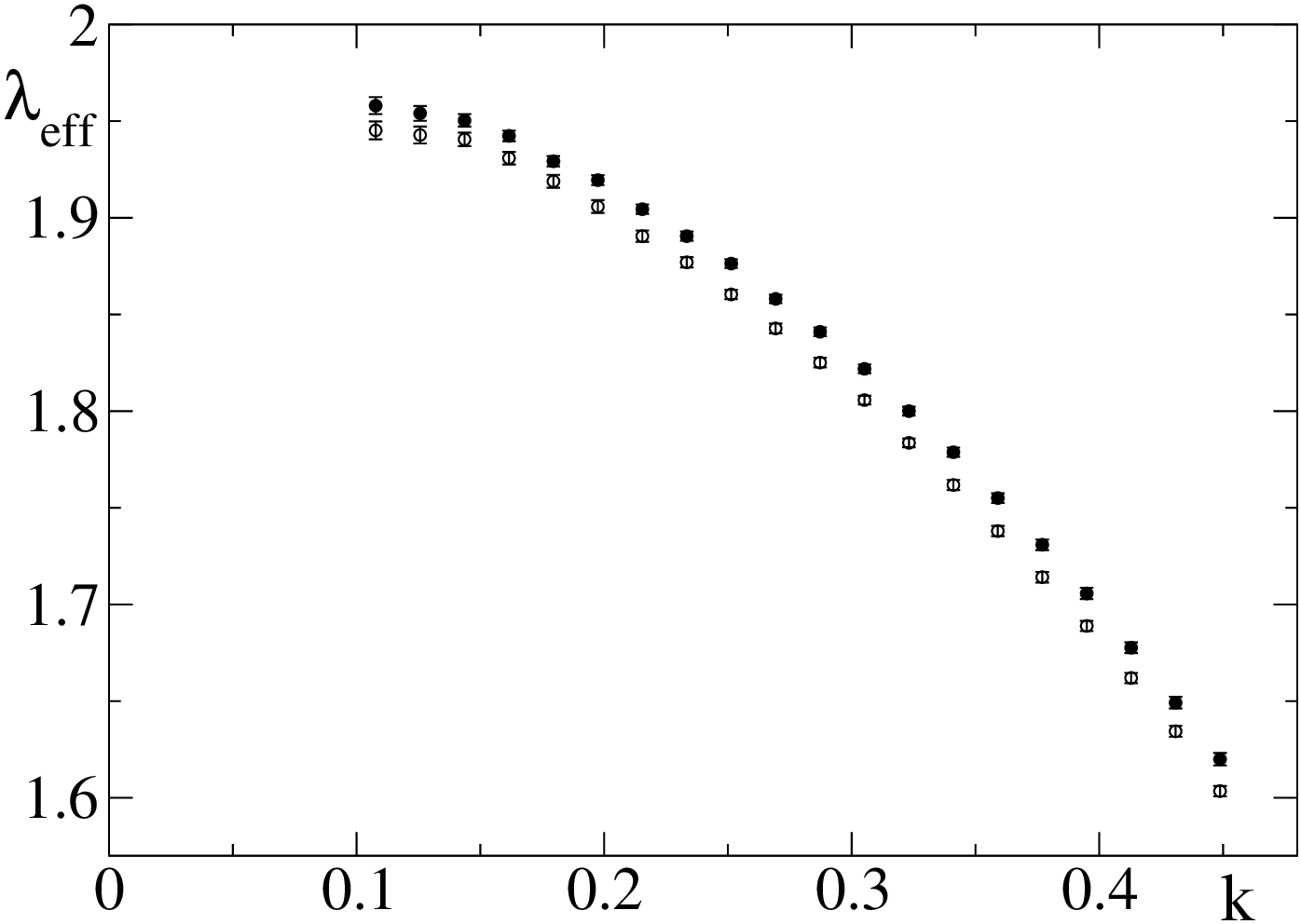}
\end{center}
\caption{The plots of the transverse effective exponent
$\lambda_{\mathrm{eff}}(k)$, evaluated from the
$\ln G_{\perp} ({\bf k})$ vs $\ln k$ fits within $[k,6k]$. The results of the $O(10)$ model at $\beta=3$
are shown by solid circles, whereas those of the $O(4)$ model at $\beta=1.1$~--- by empty circles.}
\label{lamda_compare}
\end{figure}
We have reexamined the largest-$L$ ($L=350$) and smallest-$h$ ($h=0.0003125$) data of the $O(4)$
model \cite{KMR10} at $\beta=1.1$ with an aim to evaluate the transverse effective exponent in the same way as for the $O(10)$ model.
Like in the $n=10$ case, we have verified that the thermodynamic limit at $h=+0$ is practically reached in this estimation. Besides,
we have found a surprising similarity of the $\lambda_{\mathrm{eff}}(k)$ plots, where the effective exponent in both cases
was evaluated by fitting the $G_{\perp}(\mathbf{k})$ data within $[k,6k]$ (fits within $[k,4k]$ were used in~\cite{KMR10}).
As we can see from figure~\ref{lamda_compare}, the plot for the $O(4)$ model is systematically shifted down by an almost
constant value relative to the same plot for the $O(10)$ model. This similarity might be partly caused by
the fact that the values of spontaneous magnetization are rather similar in these two cases, i.~e., $M \equiv M(+0)= 0.484475(48)$
for the $O(4)$ model at $\beta=1.1$ and $M \approx 0.467$ (see section~\ref{mag}) for the $O(10)$ model at $\beta=3$.
The overall fit to~(\ref{eq:fitansatz}) is less perfect for the $O(4)$ model as compared to the $O(10)$ model.
However, the systematical shift between two plots is very well approximated by a constant value within the statistical error
bars for the eight smallest $k$ data points in figure~\ref{lamda_compare}. It yields an estimate
$\Delta \lambda_{\perp} = \left( \lambda_{\perp} \right)_{n=10} - \left( \lambda_{\perp} \right)_{n=4}
= 0.0121(52)$, where $\left( \lambda_{\perp} \right)_{n=10}$ and $\left( \lambda_{\perp} \right)_{n=4}$
are the values of $\lambda_{\perp}$ in $n=10$ and $n=4$ cases.
According to the behavior of plots in figure~\ref{lamda_compare} and this estimation, it is quite plausible
that a transverse exponent $\lambda_{\perp}$ of the $O(10)$ model is somewhat larger than that of the $O(4)$
model. It is consistent with the idea that $2-\lambda_{\perp}$ decreases for large $n$ and tends to zero
at $n \to \infty$. This behavior is fully consistent with the predictions of~\cite{K2010}, but not so well
consistent with the standard theory, according to which $\lambda_{\perp}$ is always $2$ and, therefore,
$\Delta \lambda_{\perp} = 0$ is expected. According to our estimates
$\left( \lambda_{\perp} \right)_{n=10} =1.9723(90)$
and $\Delta \lambda_{\perp} = 0.0121(52)$, we have $\lambda_{\perp}=1.960(10)$ for $n=4$. It perfectly agrees
with our earlier estimate $\lambda_{\perp} = 1.955 \pm 0.020$~\cite{KMR10}.

\section{The ratio universality test}
\label{sec:ratio}

\begin{figure}[!b]
\begin{center}
\includegraphics[width=0.6\textwidth]{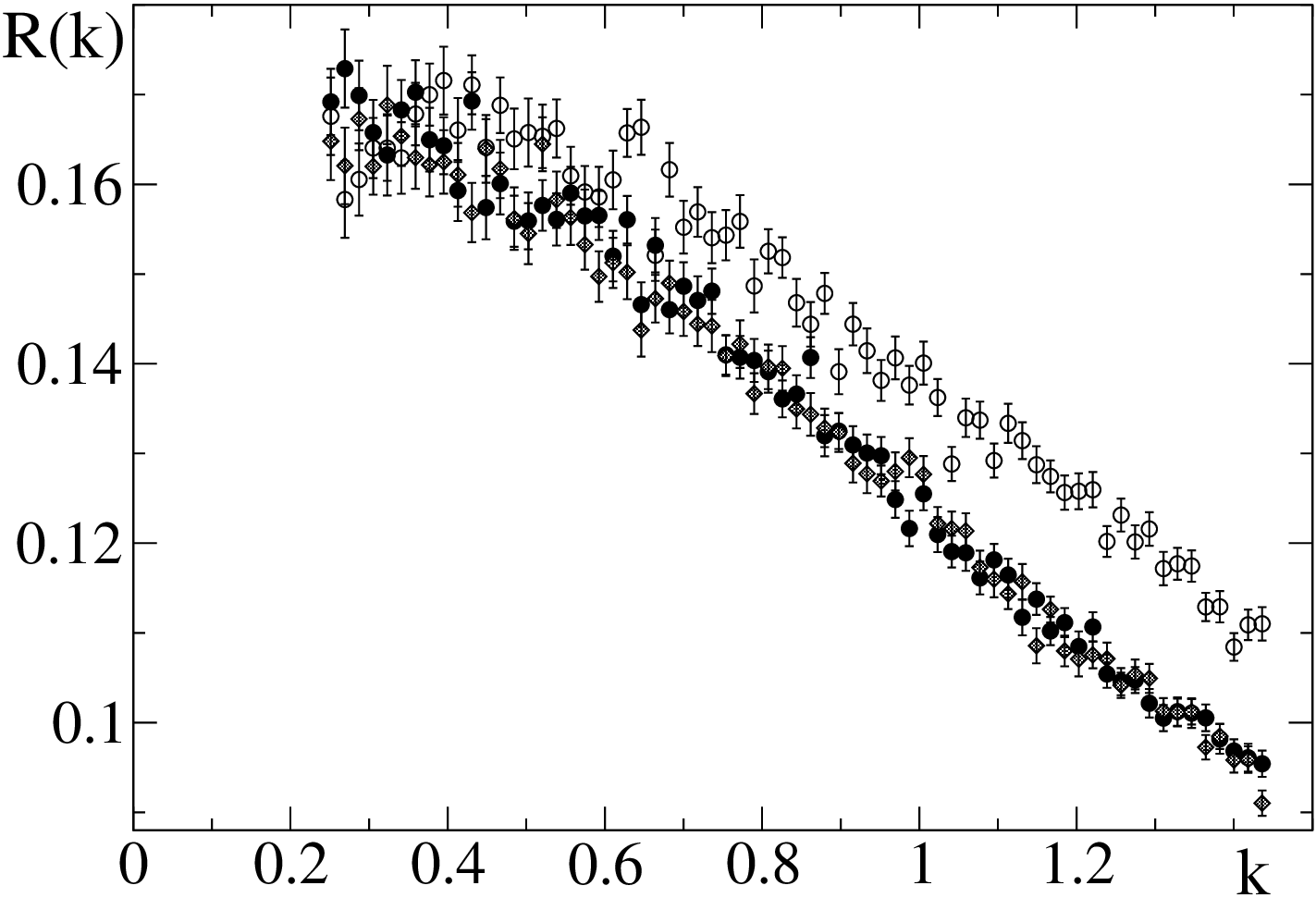}
\end{center}
\caption{The $R(k)$ plots of the 3D $O(4)$ model, evaluated from the MC data for the lattice of size $L=350$
at $\beta=1.1$ and $h=0.0003125$ (solid circles), $\beta=1.1$ and $h=0.0004375$ (diamonds), as well as at
$\beta=1.2$ and $h=0.0004375$ (empty circles).}
\label{ratio}
\end{figure}

We have extended the MC analysis of our earlier data~\cite{KMR10} for the $O(4)$ model
at two different couplings, $\beta=1.1$ and $\beta=1.2$, to test the expected (according to~\cite{K2010})
universality of the ratio $b M^2/a^2$, discussed already in the end of section~\ref{intro}.
According to~(\ref{a}), (\ref{b}) and~(\ref{eq:pred2}), the universality of $b M^2/a^2$ implies
that the quantity
\begin{equation}
 R(\mathbf{k}) = \frac{k^{-d} M^2 G_{\parallel}(\mathbf{k})}{G_{\perp}^2(\mathbf{k})}
\label{eq:R}
\end{equation}
tends to some universal constant at $k \to 0$, i.~e., $\lim\limits_{k \to 0} R(\mathbf{k}) = b M^2/a^2$.
We have tested this property by comparing the
$R(k)$ plots at $\beta=1.1$ and $\beta=1.2$, where  $R(k) \equiv R(\mid \mathbf{k} \mid)$ in the
$\langle 100 \rangle$ direction. Note that the quantities in~(\ref{eq:R}) are determined
in the thermodynamic limit at $h=+0$. We have verified that this limit is practically (within the statistical
error bars) reached within $k \geqslant k_{14}$ (where $k_{\ell}=2 \pi \ell /350$) for the largest lattice size $L=350$ and the
smallest external fields $h=0.0003125$ and $h=0.0004375$ at which  simulations were performed.
The  estimates of spontaneous magnetization obtained in~\cite{K2010}, i.e.,
$M=0.484475(48)$ at $\beta=1.1$ and $M=0.560178(40)$ at $\beta=1.2$, are used here. The calculated
plots are depicted in figure~\ref{ratio}.
The results for both $h=0.0003125$ and $h=0.0004375$ are available at $\beta=1.1$. As we can see
from figure~\ref{ratio}, the corresponding two plots of $R(k)$ (solid circles and diamonds)
lie practically on top of each other, indicating that the finite-$h$ effects are negligibly small.
The range $h \geqslant 0.0004375$ is considered for $\beta=1.2$ in~\cite{KMR10}. Fortunately,
the finite-$h$ effects at $\beta=1.2$ are similar to those at $\beta=1.1$, so that
the estimate of $R(k)$ at $h=0.0004375$ is valid at $\beta=1.2$. The corresponding plot (empty
circles) in figure~\ref{ratio}  slightly deviates from the two plots at $\beta=1.1$. However, all three
plots merge within the statistical error bars at the smallest wave vector magnitudes $k$ considered here.
This confirms the expected universality of the ratio $b M^2/a^2$.
The plot of empty circles in figure~\ref{ratio} apparently saturates at a value about $0.166$ for small
wave vectors. Taking into account the two other plots, we can judge that $0.16 < R(0) < 0.18$
most probably holds for the asymptotic value $R(0) = \lim\limits_{k \to 0} R(k)$. Thus,
we have an estimate $R(0) = 0.17 \pm 0.01$.

\section{Spontaneous magnetization}
\label{mag}

We estimated the spontaneous magnetization of the 3D $O(10)$ model at $\beta=3$
based on our magnetization data $M(h,L)$ depending on $h$ and $L$. We observed a
rather fast convergence to the thermodynamic limit, e.~g., $M(h_{\mathrm{min}},L) = 0.31658(46)$,
$0.45286(23)$, $0.470839(90)$, $0.471959(38)$, $0.472151(23)$, $0.472148(16)$
at $L=32, 64, 128, 192, 256$ and $350$, respectively.
According to this, we can take the largest-$L$ value as a good approximation for
$M(h_{\mathrm{min}})= \lim_{L \to \infty} M(h_{\mathrm{min}},L)$.
Using this method, we obtained $M(h_{\mathrm{min}}) = 0.472148(16)$,
$M(2h_{\mathrm{min}}) = 0.4742786(98)$ and $M(4h_{\mathrm{min}}) = 0.4772753(76)$.
According to~(\ref{M}), we fit these data to the ansatz $M(h) = M(+0) + a_1 h^{\rho}$
with $\rho=0.5211(69)$, evaluated from~(\ref{eq:pred3}) by inserting here
 $\lambda_{\perp} =1.9723(90)$ obtained in section~\ref{expo}. It yields
$M \equiv M(+0) = 0.467343(99)$. Assuming the standard-theoretical value $\rho=1/2$,
we obtain $M=0.467030(26)$. Fits to a refined ansatz $M(h) = M(+0) + a_1 h^{\rho} + a_2 h$
yield $M=0.46711(14)$ at $\rho=0.5211(69)$ and $M=0.46696(14)$ at $\rho=1/2$.
Since we have only three data points for $M(h)$, this can be considered as a raw estimation
yielding $M \approx 0.467$. However, this estimation is accurate enough to see that $\beta=3$
corresponds to the ordered phase with $M>0$.

\section{Conclusions}

In the actual work, the previous MC studies~\cite{KMR10,K2012} of the transverse and longitudinal
correlation functions in the 3D $O(n)$ models with $n=2$ and $n=4$ have been extended, including the
$n=10$ case (sections~\ref{sec:simu} and~\ref{expo}). It gives us an important information about the
behavior of the exponent $\lambda_{\perp}$ at large $n$. According to our MC analysis, a self-consistent
(within the standard theory) estimation
of $\lambda_{\perp}$ for $n=10$ shows a small deviation from the standard-theoretical prediction
$\lambda_{\perp}=2$, yielding $\lambda_{\perp} = 1.9723(90)$ (section~\ref{expo}). The fact that this deviation is
quite small can be well understood, since $\lambda_{\perp} \to 2$ is expected at $n \to \infty$
according to the known results for the spherical model, corresponding to this limit.
Comparing the plots of the effective transverse exponent at $n=10$ and $n=4$,
it has been stated that these plots are surprisingly similar, i.~e., only
slightly shifted with respect to each other. The estimation of this shift
suggests that the transverse exponent for $n=10$ is larger than that for $n=4$
by an amount of $\Delta \lambda_{\perp} = 0.0121(52)$ (section~\ref{expo}). It is consistent with the idea that
$2-\lambda_{\perp}$ decreases for large $n$ and tends to zero at $n \to \infty$.
We have also verified and confirmed the expected universality of
the ratio $b M^2/a^2$ for the $O(4)$ model by analyzing the correlation functions
at two different couplings, i.~e., $\beta=1.1$ and $\beta=1.2$ (section~\ref{sec:ratio}).

The actual MC results are fully consistent with the
predictions of the GFD theory~\cite{K2010} (see section~\ref{intro})
and not so well consistent with the standard
theory, according to which $\lambda_{\perp}$ is always $2$.

\section*{Acknowledgements}

This work was made possible by the facilities of the
Shared Hierarchical Academic Research Computing Network
(SHARCNET:\href{www.sharcnet.ca}{www.sharcnet.ca}).
It has been performed within the framework of the ESF Project No. 1DP/1.1.1.2.0/09/
APIA/VIAA/142, and with the financial support of this project.
R. M. acknowledges the support from the
NSERC and CRC program.

\ukrainianpart

\title{Сингулярності голдстоунівських мод в $O(n)$ моделях}

\author{Я. Каупузс\refaddr{label1,label2},
Р.В.Н. Мельнік\refaddr{label3}, Я. Рімсанс\refaddr{label1,label2}}

\addresses{
\addr{label1} Інститут математики та комп'ютерних наук, Університет
Латвії, \\ бульвар Я. Райніса, 29, LV--1459 Рига, Латвія\addr{label2}
Інститут математичних наук та інформаційних
технологій, Університет м. Лієпая, \\  вул. Лієла, 14, LV--3401 Лієпая, Латвія
\addr{label3}Університет ім. Вільфреда Лорьє,
Ватерлоо, Онтаріо, Канада, N2L 3C5 }

\makeukrtitle

\begin{abstract}
\tolerance=3000%
У тривимірних $O(n)$ моделях з $n=2,4,10$ здійснено аналіз методом
Монте Карло (МК) сингулярностей голдстоунівських мод для поперечної
і поздовжньої кореляційних функцій, які поводять себе як  $G_{\perp}
({\bf k}) \simeq a k^{-\lambda_{\perp}}$ і $G_{\parallel} ({\bf k})
\simeq b k^{-\lambda_{\parallel}}$ у впорядкованій фазі при $k \to
0$. Нашою метою є перевірити цікаві теоретичні передбаченя, згідно
яких індекси $\lambda_{\perp}$ і $\lambda_{\parallel}$ є
нетривіальними ($3/2<\lambda_{\perp}<2$ і $0<\lambda_{\parallel}<1$
у трьох вимірах) і коефіцієнт $b M^2/a^2$ (де $M$ є спонтанною
намагніченістю) є універсальний. Тривіальні стандартні теоретичні
значення є $\lambda_{\perp}=2$ і $\lambda_{\parallel}=1$. Наш
попередній МК аналіз дає $\lambda_{\perp}=1.955 \pm 0.020$ і
$\lambda_{\parallel}$ приблизно рівне $0.9$ для $O(4)$ моделі.
Недавня МК оцінка $\lambda_{\parallel}$, яка допускає поправки для
скейлінга стандартної моделі, дає $\lambda_{\parallel} = 0.69 \pm
0.10$ для $O(2)$ моделі. Тепер ми здійснили подібну МК оцінку
 для $O(10)$ моделі, яка дає $\lambda_{\perp} = 1.9723(90)$. Ми
 побачили, що графік ефективного поперечного індекса  для $O(4)$
 моделі є систематично зсунутий вниз по відношенню до графіка для
 $O(10)$ моделі на $\Delta
\lambda_{\perp} = 0.0121(52)$. Це узгоджується з думкою, що
$2-\lambda_{\perp}$ зменшується для великих $n$ і прямує до нуля при
$n \to \infty$. Ми також перевірили і підтвердили очікувану
універсальність $b M^2/a^2$ для  $O(4)$ моделі, для якої було
здійснено симуляції при двох різних температурах.
\keywords моделювання Монте Карло, $n$-компонентні векторні моделі,
кореляційні функції, сингулярності голдстоунівських мод
\end{abstract}
\end{document}